\documentclass[10pt,letterpaper,twocolumn]{article} 

\usepackage{ol2}
\usepackage{amsmath,amssymb}

\begin{document}

\twocolumn[ 


\title{\LARGE Active optics for high dynamic variable curvature mirrors}

\author{ \small Emmanuel Hugot$^{*}$, Marc Ferrari, G\'erard R.Lemaitre, Fabrice Madec, S\'ebastien Vives, Elodie Chardin, David Le Mignant and Jean-Gabriel Cuby}
\address{$^1$ Aix Marseille University, CNRS, Laboratoire d'Astrophysique de Marseille, 13388 Marseille Cedex 13, France}

\begin{abstract}
    Variable curvature mirrors of large amplitude are designed using finite element analysis. The specific case studied reaches at least a 800$\mu$m sag with an optical quality better than $\lambda/5$ over a 120mm clear aperture. We highlight the geometrical non linearity and the plasticity effect.
\end{abstract}


] 

\section{Variable Curvature Mirrors}
    Following the work of Lemaitre on Active Optics in 1976 \cite{lemaitre-vcm}, high dynamic variable curvature mirrors (VCMs) have been developed by Ferrari in 1994 \cite{ferrari-vcm} for the delay lines of the very large telescope interferometer (VLTI). These metallic mirrors are bent spherically by a pressurization on the back of a 300$\mu$m thin meniscus, allowing a stroke of 380$\mu$m on a 16mm diameter mirror with a pressure of 0.8MPa, resulting in a variable curvature from 2800mm$^{-1}$ to 84mm$^{-1}$.

\begin{figure}
            \begin{center}
                \includegraphics[width=8.4cm]{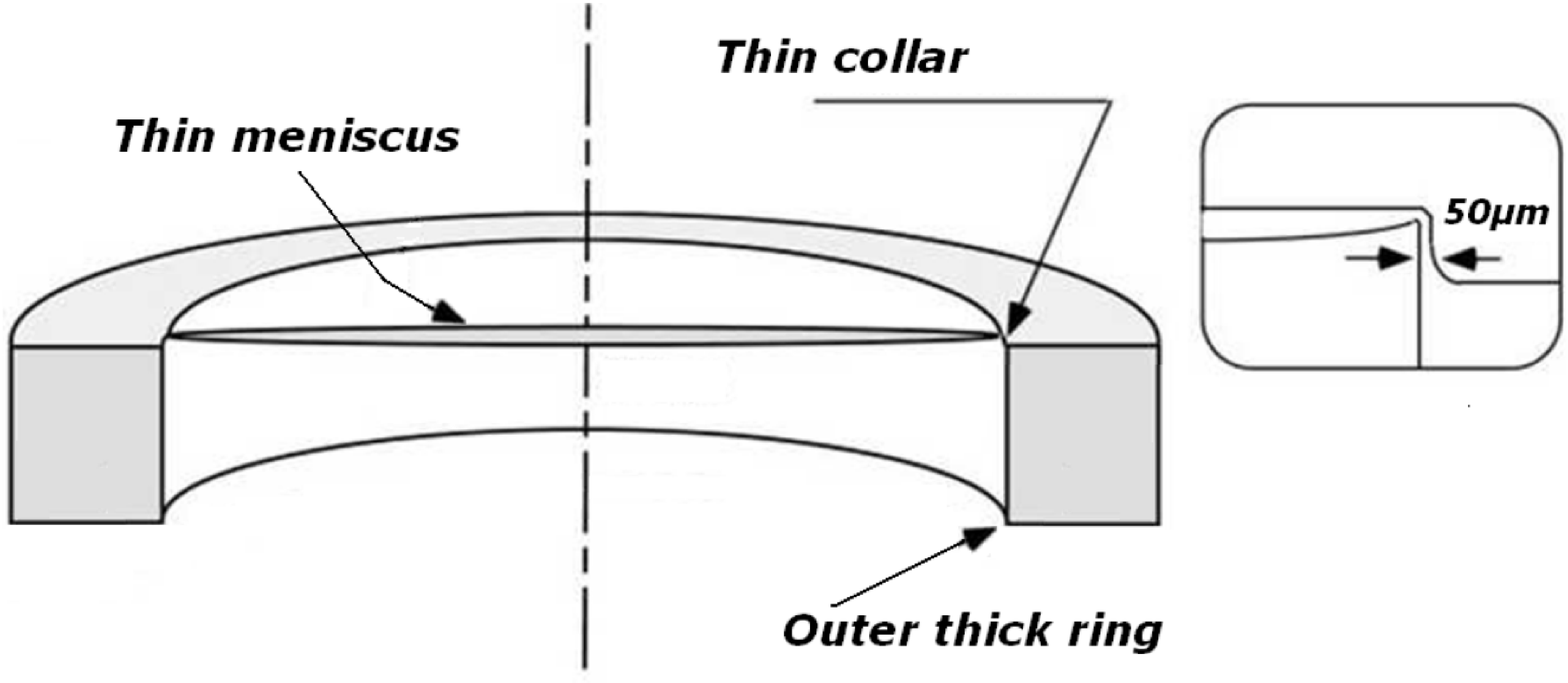}
                \caption{\label{fig:VCM-profile}
                {\small Variable curvature mirror profile and zoom on the thin collar.}
                \vspace{-0.8cm}}
            \end{center}
        \end{figure}

    A variable thickness distribution (VTD) is machined on the meniscus in order to ensure a spherical deflection with an optical quality better than $\lambda/4$ over the full range. This VTD is obtained from the elasticity theory \cite{Timo}. The variable thickness distribution is defined in the linear case by $t(\rho) = t_0(1-\rho^2)^{1/3}$ \cite{lemaitre-book}, where $t(\rho)$ is the variable thickness, $\rho$ is the normalised radius of the surface and $t_0$ is the central thickness. The design of the VCM uses a \emph{holosteric} solution where the edge of the central active meniscus is linked with an outer rigid ring via a very thin collar, and an air pressure is applied on the back side of the meniscus. This simplifies the mounting and avoids air leak by having the mirror and its outer ring in the same sample (Fig.\ref{fig:VCM-profile}).

    In the case of VLTI VCMs, the maximal sag is larger than 1.25 times the central thickness of the meniscus. Keeping the optical quality requires taking into account the large displacement non linearity that appears when the sag is higher than 1/3 of the meniscus thickness. The deflection of the optical surface is degraded by a spherical aberration that increases with the sag of the vertex. To compensate this effect, the VTD is adjusted with the non linear model and the evolution of the spherical aberration is minimized on 85\% of the diameter, over the full range of deflection.

    To further study the performance of the VCMs, we investigate the design for achieving a sag of 1mm with the same optical quality. The aim of this article is the definition of the VCM's geometry based on analytical study and finite element analysis (FEA). The next key point will be the metrology for such an active system. For that, studies are under going on how to measure the two main parameters: the optical quality at any curvature, and the curvature measurement linked to the sag.  This will be emphasized in a future article.

\section{Analytic Definition of VCM's Geometry}

    For analytical study, we consider the same pressure limit, and a thickness higher than 3 times the maximal sag to avoid the large displacement non linearity. We also state that, to avoid edges effects, the clear aperture of the mirror must be 85\% of the total diameter of the mirror as for VLTI VCMs. We go further by considering a convex meniscus of radius equal to 2800mm in its stress free state.

    These assumptions drive the VCMs parameters. Let's define $a$ as the radius of the meniscus, $q_{max}$ as the maximal load applied, $E$ and $\nu$ as the Young's modulus and Poisson's ratio of the material. From the theory of circular plates of constant thickness simply supported on the edge, the maximal sag $w_{max}$ at center of the meniscus is given by (Timoshenko \cite{Timop57}):
        \begin{equation}
            w_{max} = \dfrac{(5+\nu)q_{max}a^4}{64(1+\nu)D} \quad,\quad   D = \dfrac{Et_0^3}{12(1-\nu^2)}
        \end{equation}
    Considering that $t_0 \geq 3w_{max}$ in Eq.\ref{equ:w-vcm}, the radius $a$ varies linearly with the maximal sag $w_{max}$:
        \begin{equation}
            a \geq 2 \left[ \dfrac{(1 + \nu)}{(5 + \nu)(1 - \nu^2)} \dfrac{9E}{q_{max}} \right]^{1/4} \times w_{max}
        \end{equation}
    A better accurate approach is also given by Pichler \cite{Pichler} by considering a circular plate of non uniform thickness simply supported on the edge. In our case the thickness profile can be roughly approximated by $t(\rho) = t_0 e^{-x^2/2}$. This approximation allows us to use the results given in \cite{Timop298}. The maximal sag at center is given by:
        \begin{equation}\label{equ:w-vcm}
            w_{max} = \dfrac{1.2 (1-\nu^2) a^4 q_{max}}{Et_0^3}
        \end{equation}
    with $t_0 \geq 3w_{max}$ we then get:
        \begin{equation}\label{equ:a-vcm}
            a \geq \left[ \dfrac{27E}{1.2 (1-\nu^2) q_{max}} \right]^{1/4} \times w_{max}
        \end{equation}
    Assuming that the deflection is a pure sphere, the sag between the vertex and the edge of the clear aperture is $w_{0.85} = 0.85^2w_{max}$.

    From that we are able to define the mirror's basic parameters.  Considering a metallic VCM made in AISI420 with $E$ = 215000MPa and $\nu$ = 0.305 and a maximal pressure $q_{max} = 0.8$MPa, Eq.\ref{equ:a-vcm} gives $a = 50.81w_{max}$.  For a sag $w_{0.85} = 1.0mm$, we have $w_{max} = 1.384mm$, leading to a total diameter of the mirror of about 140.00mm and a thickness of 4.15mm. That means we will consider an F/10 convex meniscus. From this preliminary definition, the FEA will allow to quantify the stress within the material, adjust the geometrical parameters and compute the optical quality of the deflection obtained. Even if the large displacement non linearity is avoided, we will see that a geometrical non linearity must be taken into account for convex VCMs. We also study the case of a plastic deformation of the collar.

\section{Finite Element Analysis}
    Analytical calculations are helpful for a preliminary definition of the blank's geometry but do not take into account several mechanical details in the final design of the mirror. To optimize this preliminary geometry, we perform FEA of this Active Optics device using Marc/Mentat software \cite{marcmentat}. The axisymmetric model we compute is made of 43,000 quadrangles, 44,000 nodes, and the sampling on the optical radius is of 500 points. In this case, we fix the thickness of the collar at 450$\mu$m to keep the existing VLTI VCMs thickness ratio.
    \subsection{Preliminary Results}
        The first result produced by the FEA is that the deflection obtained with the geometrical parameters previously defined is lower than expected. The maximal displacement of the vertex is lower than 1.3mm instead of 1.384mm. This difference is due to the thin collar at the edge which produces a "semi built-in" effect, equivalent to a rigidity increase, depending to the ratio between the collar and the meniscus thickness. Furthermore, this collar adds a non negligible amount of error compare to the spherical deflection expected, error which must be compensated by a modification of the VTD. Moreover, a geometrical non linearity appears when the deflection of the vertex is around 1/10 of the maximal thickness of the meniscus (see next section). Finally the maximal stress level is obtained on the collar, with a value around 1400 MPa. For standard AISI420, the elastic limit is between 600MPa and 800MPa, but can be increased up to 1400MPa with a specific thermal treatment. That means we must take into account a possible plastic deformation of the material in the case of standard AISI.

    \subsection{Geometrical Non Linearity - Modification of the VTD}
\begin{figure}
        \begin{center}
                \includegraphics[width=8.4cm]{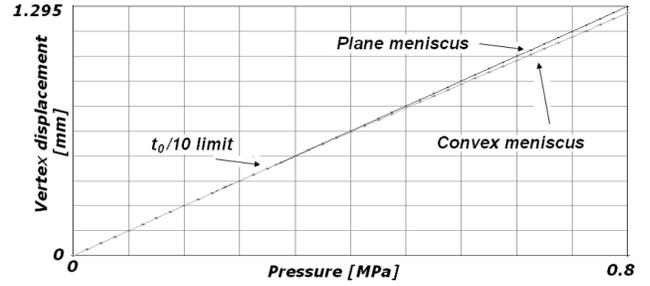}
                \caption{\label{fig:VCM-NL}
                {\small The vertex displacement curve shows that a geometrical non linearity appears when the sag is equal to 1/10 of the central thickness.}\vspace{-0.8cm}}
        \end{center}
        \end{figure}

        Considering an F/10 meniscus, even if the F-number is quite large, introduces a shell effect in the deflection non negligible for the sag we want to reach. In fact, a geometrical non linearity appears when the sag is higher than 1/10 of the meniscus maximal thickness. This effect is highlighted by the vertex displacement plotted on Fig. \ref{fig:VCM-NL}, which is linear for a plane meniscus and non linear for an F/10 meniscus. This slight deviation will introduce a surface error during the bending, that can be compensated by a modification of the VTD.

        In order to optimize the VTD, we use a specific iterative routine on the geometrical parameters of the FEA model, with a criterion based on the optical quality of the mechanical deflection. The RMS surface error is computed from the displacement of the points by interpolating the best fit sphere of the curve with a root mean square minimisation.

\begin{figure}
            \begin{center}
                \includegraphics[width=8.4cm]{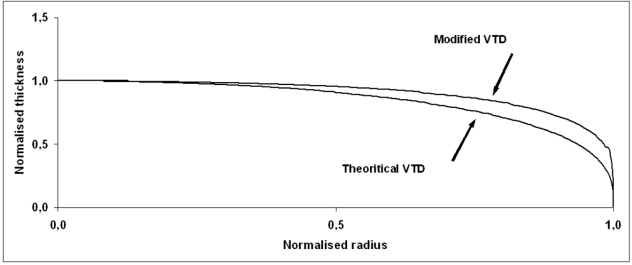}
                \caption{\label{fig:VCM-modif-VTD} {\small A modification of the VTD allows compensating for the non linear behavior. This modification is obtained by a finite element optimization.} \vspace{-0.8cm}
                }
            \end{center}
        \end{figure}

        The final modification is shown on Fig.\ref{fig:VCM-modif-VTD}. The VTD has been optimized in order to equilibrate the surface error over the full curvature range. The effect of the non linearity on the surface error is highlighted on Fig.\ref{fig:VCM-RMS-curves02} and on the top of Tab.\ref{tab:residue RMS}, where the plain curve draws the evolution of the RMS surface error versus deflection on the clear aperture (85\% of the total deflection). This result shows that the surface error is well equilibrated all over the pressure variation in order to be lower than $\lambda/5$ up to 940$\mu$m of sag, 5\% lower than the 1mm expected.

    \subsection{Plasticity}
        In the case of a standard AISI, the stress level in the collar is higher than the elastic limit. That means the material enters the plastic deformation domain and will get a permanent deformation after removal of the loads. Several questions appear. We need to know what is the amplitude of the permanent deformation after removal of the load, if this deformation is spherical and with which optical quality and finally what is the behavior of the mirror after plastic deformation.

        In a preliminary approach, we consider in the simulations an AISI420 with an elastic limit equal to 800MPa. We use  the Von Mises criterion considering an average of the stress in the principal directions. FEA shows that this stress value is obtained for a pressure of 0.48MPa and 560$\mu$m of sag on the clear aperture.

\begin{figure}[h]
            \begin{center}
                \includegraphics[width=8.4cm]{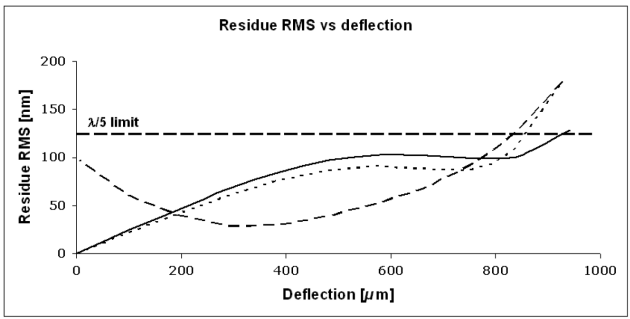}
                \caption{\label{fig:VCM-RMS-curves02}
                {\small Plot of RMS residual versus the sag on the clear aperture. Plain curve: no plasticity. Dotted curve: before plastic deformation. Dashed curve: after plastic deformation.} \vspace{-0.8cm}}
            \end{center}
        \end{figure}

        Fig.\ref{fig:VCM-RMS-curves02} plots the RMS surface error versus the sag. The dotted curve shows the change of the optical quality, starting from a stress free mirror. The behavior is slightly different compare to the case without plasticity (plain curve) and is lower than $\lambda/3$ for a pressure of 0.8MPa. After plastic deformation, the change of the optical quality is given by the dashed curve. The permanent deformation has an amplitude of 5.3$\mu$m, the shape is spherical with a residue lower than 100nm RMS. A minimum is visible for 330$\mu$m of sag and up to 825$\mu$m of deflection on the clear aperture, the residue is lower than 125nm RMS . After the pre-stressing, meaning the plastic deformation of the mirror, the residue will follow the dashed curve. From these data we can conclude that the VTD modification is also usable in the case of a plastic deformation of the collar on a reduced dynamic range. Tab.\ref{tab:residue RMS} gives some numerical values corresponding to Fig.\ref{fig:VCM-RMS-curves02}.

\section{Conclusion and future work}
    We designed a convex variable curvature mirror able to provide a large sag variation on a spherical surface with a 120mm clear aperture, with an optical quality better than $\lambda$/5 up to 950$\mu$m of sag. A geometrical non linearity appearing during the deflection introduces a variation in the global shape of the deflection. To compensate for that, we optimized the variable thickness distribution in order to equilibrate the residual error over the full curvature range. In the case of a plastic deformation of the VCM's collar, the optical quality evolution is slightly different. A permanent deformation remains after removal of the loads and the $\lambda$/5 quality is reached from 5.3$\mu$m to 825$\mu$m of sag on the clear aperture. Two prototypes are under fabrication to compare the real behavior of the mirror and the simulation data, made in thermal-treated and standard AISI420. Metrology, performances and comparison with the FEA model of these Active Optics devices will be presented in a future article. This work is supported by The French Agence Nationale de la Recherche (ANR) program 06-BLAN-0191 and the European Community (Framework Programme 7, E-ELT Preparation, contract No INFRA-2.2.1.28)
 \begin{table}
        \begin{center}
        \caption{\label{tab:residue RMS} {\small Deflection amplitude and RMS residual with respect to a sphere in the case without plasticity, before plastic deformation and after plastic deformation.}}
        \begin{tabular}{l c c c c c}
            \hline
            Pressure [Mpa]      &   0   &   0.24	&  0.48    &   0.64    &   0.8   \\
            \hline
            \emph{No plasticity}	                                                         \\
            Deflection [$\mu$m] &   0   &   292.2	& 577.4   &   761.9 & 940.8      \\
            Residual [nm]       &   0   &   67.8	&  103.1  &    99.0	&  128.8     \\
            \hline
            \emph{Before plasticity}												         \\
            Deflection [$\mu$m] &   0   &   285.6   &  564.7    & 746.4	& 927.2      \\
            Residual [nm]       &   0   &   59.7    &   90.7    &  87.2	&  179.4     \\
            \hline
            \emph{After plasticity}												         \\
            Deflection [$\mu$m] &   5.3 &   291.3	& 661.9   & 751.7	&  927.2     \\
            Residual [nm]       &   97.2&   28.7	&  52.2    &  92.3	&   179.4    \\
            \hline 
        \end{tabular}
        \end{center}
        \end{table}







\end{document}